# О РАЗБИЕНИЯХ ТИПА ПЕНРОУЗА ДВУМЕРНОЙ СФЕРЫ МОДЕЛИРУЮЩИХ КВАЗИКРИСТАЛЛЫ

**Александр С. Прохода, к.ф.-м.н.**

Днепр, Украина

Решается задача о построении предельного ряда разбиений типа Пенроуза двумерной сферы, что позволяет моделировать квазикристаллы обладающие точечной икосаэдральной группой симметрии $I_h$. Приведены модели многогранников число граней, у которых F > 5000. Исходя из определенных сферических изоэдральных многогранников, описан рецепт построения сферических многогранников Платона, Архимеда, Каталана и Джонсона. Установлены границы хроматического числа в пространстве $S^2$.

## 1. Введение

Известное квазипериодическое разбиение Пенроуза (Р3) евклидовой плоскости двумя ромбами – ромбами Пенроуза (с острыми углами 72° и 36°), моделирующее двумерный квазикристалл, обладает двумя замечательными свойствами, отличающими его от разбиений евклидовой плоскости на планигоны и плоские изогоны, которые в свою очередь моделируют двумерные периодические (классические) кристаллы. Первое свойство заключается в следующем: фундаментальная область группы симметрий разбиения не может быть выбрана только в виде одного многоугольника. Группа симметрии разбиения Пенроуза (Р3) обладает фундаментальной областью состоящей из двух треугольников с углами (108°, 36°, 36°) и (72°, 72°, 36°). Эти треугольники называются треугольниками Робинсона и получаются путем "разрезания" ромбов Пенроуза по соответствующим диагоналям.

______________________

E-mail: a-prokhoda@mail.ru

Треугольник с углами (108°, 36°, 36°) подходит также под определение треугольника Шварца, т.е. такого треугольника, у которого углы являются рациональными частями π. Треугольник же с углами (72°, 72°, 36°) обладает тем свойством, что его углы являются целыми частями от 360°. Поэтому символом данного треугольника является тройка чисел (5, 5, 10). Треугольник (5, 5, 10) исследовал Делоне в работе [1]. Делоне установил, что данным треугольником невозможно произвести разбиение "грань в грань" евклидовой плоскости. Условие, что углы многоугольника являются целыми частями 2π, есть необходимым, но недостаточным условием существования разбиения плоскости на планигоны. Делоне в [1] нашел все возможные планигоны. Второе свойство разбиения Пенроуза (P3) заключается в том, что оно не является изогональным, так как в вершинах этого разбиения сходится различное число многоугольников. При этом множество вершин разбиения Пенроуза является правильной ($r$, $R$)-системой точек Делоне (за исключением условия транзитивности). Все возможные варианты схождений тайлов около вершин в разбиении Пенроуза (P3) описаны в работе [2], которые определяются локальными правилами самосборки.

Таким образом, можно дать определение разбиения типа Пенроуза поверхности (замкнутой или незамкнутой, гладкой или кусочно-гладкой) на криволинейные многоугольники.

**Определение 1.** Пусть дана двумерная геометрия $\Gamma = (X, G)$, где $X$ – евклидова плоскость $E^2$ или двумерная сфера $S^2$, а $G$ – группа, действующая на $X$, тогда разбиение носителя геометрии $X$ называется разбиением типа Пенроуза, если оно обладает следующими свойствами: 1) разбиение осуществляется при помощи неконгруэнтных многоугольников (разбиение не является изоэдральным); 2) разбиение не является изогональным; 3) разбиение обладает ориентационным дальним порядком.

Поскольку существуют различные типы дальних порядков, приведем определение дальнего порядка.

**Определение 2.** Говорят, что кристаллическая структура пространства обладает дальним порядком, если, где бы ни была расположена точка пространства можно, указать алгоритм, позволяющий в конечное число шагов определить, расположен или не расположен элемент кристаллической структуры (атом, молекула и т. д.) в этой точке.

Наличие дальнего порядка с теоретико-групповой точки зрения означает, что группа симметрий кристалла должна содержать бесконечную подгруппу, которая действует в пространстве и определяет этот порядок. В случае обычных кристаллов такой подгруппой является подгруппа параллельных переносов, которая представляет собой свободную абелеву группу, ранг которой (т.е. максимальное число независимых элементов) равен размерности пространства. Решеткой в $n$-мерном пространстве называется множество точек с целочисленными аффинными координатами. Другими словами решеткой в $n$-мерном пространстве называется, свободная абелева группа ранга $n$. В случае физического евклидова пространства $E^n$ (при $n$ равным 2 или 3) решетка не может обладать поворотными симметриями пятого и выше шестого порядка. Однако квазикристаллы обладают некоторыми указанными симметриями.

В данной работе выполнены разбиения типа Пенроуза носителя геометрии $\Gamma = (S^2, O(3))$, т.е. сферы $S^2$ на криволинейные многоугольники. Что позволило построить и исследовать самостоятельный ряд сферических многогранников обладающих точечной икосаэдральной группой. При этом была установлена верхняя граница значения хроматического числа в пространстве $S^2$. Однако прежде чем переходить к результатам данного исследования необходимо условиться о соответствующей терминологии.

Задачи изогональных и изоэдральных разбиений евклидовой плоскости были решены в классических работах Федорова [3], Шубникова [4], а также Делоне [1]. Согласно Федорову [5, 6] изогоном называется многогранник, в каждой вершине которого сходится по одному и тому же числу ребер или граней. Типическим изогоном называется многогранник с равными многогранными углами (гоноэдрами). Термин гоноэдр ввел в обиход Федоров в [5] вместо термина "многогранный угол", например тригоноэдр – трехгранный угол. Изоэдром называется многогранник у которого все грани одного и того же наименования. Подтипический изоэдр получается из типического изогона так. Опишем сферу около типического изогона, проведем касательные к сфере плоскости через вершины; эти плоскости образуют подтипический изоэдр. В подтипическом изоэдре все грани имеют не только одно наименование, но они все равны (конгруэнтны) между собой. В двумерном случае Федоров называл многоугольник планигоном, если существует нормальное разбиение плоскости на эти многоугольники. И это разбиение правильное в смысле Федорова, т.е. чтобы всякий из этих многоугольников был окружен всеми другими до бесконечности точно так же, как всякий другой из них. Разбиение называется правильным, если группа симметрий действует на нем транзитивно. Другими словами, какие бы небыли два многоугольника разбиения существует движение, принадлежащее группе симметрий переводящее один многоугольник в другой. Согласно терминологии, принятой в [7, 8] правильное разбиение плоскости на планигоны называется изоэдральным разбиением, дуальное же к нему разбиение называется изогональным разбиением. Изогональные разбиения плоскости описаны Шубниковым в [4]. Изоэдральные разбиения плоскости (дуальные к изогональным разбиениям) были получены Делоне в [1]. Лавес в [9] так же рассмотрел задачу о разбиении плоскости на планигоны.

Если у изогона все многогранные углы равны, то у дуального изоэдра все грани одного и того же наименования, и можно выбором вершин добиться, чтобы все грани его были равны. Способ построения изоэдра по изогону описан Федоровым в [5]. Если изогон нетипический, т.е. в каждой его вершине сходится одно и то же число ребер, но гоноэдры не все равны, то дуальный к нему изоэдр может иметь неравные грани, причем одного и того же наименования. Шубников в [4] доказал теорему, из которой следует, что в изогоне не может быть более пяти различных по наименованию граней. Таким образом, изогонами являются многогранники Архимеда, а изоэдрами – многогранники Каталана. Следует отметить, что планигон можно считать плоским изоэдром. Разбиение поверхности $M$ евклидового пространства $E^3$ на многоугольники называется моноэдральным, если все его многоугольники попарно конгруэнтны. В правильном изоэдральном разбиении все многоугольники разбиения попарно конгруэнтны, т.е. оно является моноэдральным разбиением. Обратное же утверждение не всегда верно.

Мадисон в работе [10] развивая идеи Феликса Клейна [11] пишет: в частности, изучение апериодического разбиения евклидовой плоскости мы вправе заменить проективным исследованием некоторого его образа на поверхности второго порядка (например, на сфере). Вспоминая при этом, что геодезическая (геодезическая линия) – это обобщение понятия прямой линии в искривленных пространствах. Под расстоянием, измеренным по поверхности (ненулевой кривизны), понимается длина дуги соответствующей геодезической линии.

Напомним, что многогранник, у которого все грани правильные многоугольники и все многогранные углы равны, называется однородным, также в однородных многогранниках каждую вершину окружают многоугольники в одном и том же порядке. Любой однородный многогранник можно поместить внутри сферы таким образом, что его оси симметрии пройдут через центр сферы. Спроектировав затем из центра на поверхность сферы ребра многогранника, мы получим сеть, состоящую из дуг больших окружностей сферы. Эта сеть разбивает сферу на сферические многоугольники, каждый из которых соответствует одной грани многогранника. Плоскости симметрии многогранника добавят к разбиению новые дуги, так что если исходный многогранник, к примеру, был

Платоновым телом, то с учетом новых дуг поверхность сферы будет разделена на сферические треугольники – по четыре для каждого ребра. Эти сферические треугольники получили название треугольников Мёбиуса, по имени впервые рассмотревшего их математика в 1849 г. (см. [12]). Вспомним, что в геометрии имеется одно семейство треугольников именуемых треугольниками Шварца [13]. Причем в случае треугольников Шварца путем отражений от их сторон разбиения сферы осуществляются с наложениями одних треугольников на другие. Каждый треугольник Шварца на сфере определяет конечную группу. Треугольник Шварца представляется тремя рациональными числами ($α_1$, $α_2$, $α_3$), каждое из которых задает внутренний угол при вершине треугольника. Если эти числа целые, то треугольник будет треугольником Мёбиуса, и он соответствует мозаике без перекрытий. На сфере имеется три треугольника Мёбиуса: (3, 3, 2) – тетраэдральная симметрия; (4, 3, 2) – октаэдральная симметрия; (5, 3, 2) – икосаэдральная симметрия, а также одно однопараметрическое семейство ($d$, 2, 2) – диэдральная симметрия, где $d$ натуральное число. Отметим, что в икосаэдральных треугольниках Шварца максимальным разрешенным числителем может быть число 5. В геометрии $d$ – угольным осоэдром (диэдральная группа симметрии), называется такая мозаика из двуугольников на сфере, что каждый такой двуугольник имеет две общие вершины (расположенных в противоположных точках сферы, т.е. на ее полюсах) с другими двуугольниками. Многомерный аналог осоэдра называется осотопом. Каждый $d$ – угольный осоэдр связан с соответствующим аксиальным квазикристаллом и может являться его сферической моделью.

Если выбрать какой-либо порядок цветов и раскрасить грани примыкающие к некоторой вершине многогранника в этом порядке по ходу часовой стрелки, то энантиоморфной раскраской будет обратная – в том же порядке, но против часовой стрелки. Раскраска построенных в данной работе моделей осуществлялась по основному принципу раскраски карт. Это означает, что грани многогранника, имеющие общее ребро, должны быть окрашены в разные цвета.

## 2. Методика построения и анализ предельного ряда

Как топологическое пространство сфера является родоначальником двух серий замкнутых двумерных многообразий – сфер с $g$-ручками, где $g$ – род многообразия и проективных плоскостей $RP^2$, где $RP^2$ – результат отождествления всех диаметрально противоположных точек сферы (в случае неориентированных поверхностей к сфере необходимо добавить $q$-пленок Мёбиуса). Знаменитая формула $\Phi = gT^2 \# qRP^2,\ \Phi \# S^2 = \Phi$ (здесь $\#$ – связная сумма многообразий) несет столь же богатую информацию, как и формула $e^{2\pi i} = -1$ или формула Эйнштейна $E = mc^2$. Формула $\Phi \# S^2 = \Phi$ означает, что в случае сферы ее род $g$ равен нулю, а также $q = 0$, то есть сфера – ориентированная поверхность рода ноль. Топологическим инвариантом сферы является ее эйлерова характеристика $\chi(S^2) = 2$. Что же касается эйлеровой характеристики $\chi(\Phi)$ любой замкнутой поверхности, то она связана с родом следующей формулой $\chi = 2 - 2g - q$. Числа $\chi$, $g$, $q$ являются топологическими инвариантами двумерных многообразий, они же являются инвариантами фундаментальной группы поверхности.

Понятие рода поверхности тесно связано с понятием связности поверхности. Для замкнутых поверхностей имеем следующее определение.

**Определение 3.** Замкнутая поверхность $S$ называется $h$-связной, если на ней можно провести $h - 1$ замкнутых кривых, не разбивающих поверхность на части (которые можно отделить друг от друга), но всякая система, состоящая из $h$ подобных кривых разбивает поверхность.

**Примечание.** Можно подчинить эти кривые к еще одному условию – потребовать, чтобы они проходили через определенную произвольную точку поверхности. Таким образом, можно получить "каноническую" систему разрезов поверхности пригодную для многих целей.

Например, сфера $S^2$, очевидно, односвязна: $h = 1$, так как любая замкнутая кривая ее разбивает. Можно доказать, что связность $h$ и род $g$ удовлетворяют соотношению: $h = 2g + 1$, а ейлерова характеристика поверхности $S$ задается формулой: $\chi(S) = V - E + F = 3 - h$. Так как $h = 2g + 1$, то получаем, что

$$\chi(S) = V - E + F = 2 - 2g.$$

Так как эйлерова характеристика сферы равна 2, то род сферы равен 0. У тора $h = 3$, следовательно, род тора равен 1. Для поверхности нечетной связности можно доказать, что на замкнутой поверхности связности $h = 2g + 1$ имеется $g$ и не более чем $g$ замкнутых не пересекающих друг друга кривых, не разбивающих данную поверхность. Например, для тора $h = 3$, а $g = 1$, $\chi(T^2) = 0$. Род сферы равен нуль, т.к. из теореме Эйлера о выпуклых многогранниках следует, что $\chi(S^2) = V - E + F = 2$.

В теории многообразий (которые можно отнести к комбинаторной топологии) особую роль играют разбиения многообразий на клетки, которые составляют так называемый клеточный комплекс. Его клетки можно рассматривать как отдельно, так и как элементы разбиения, которые получаются путем операций склеивания, приклеивания и т.д. Приведем определение клетки.

**Определение 4.** Клетка – это подмножество $K \subset X$ произвольного хаусдорфова пространства $X$, гомеоморфное открытому диску некоторой размерности. Грани это 2-клетки, ребра – 1-клетки, а вершины – 0-клетки. Клеточным разбиением называется такое разбиение многообразия $X$ на клетки, у которого граница любой клетки содержится в объединении клеток меньшей размерности.

Теперь мы можем привести следующее определение.

**Определение 5.** Эйлеровой характеристикой пространства $X$ имеющего конечное клеточное разбиение, определяется как число клеток четной размерности минус число клеток нечетной размерности.

Для сферы $S^2$ это и есть формула Эйлера.

Приведем также понятие клеточного комплекса, которое введем по индукции. Клеточный комплекс размерности ноль – это конечный набор точек. Клеточный комплекс размерности один получается из него приклеиванием нескольких одномерных клеток (дуг) по отображениям их краев (пар точек). Клеточный комплекс размерности два строится путем приклеивания к одномерному комплексу двумерных клеток (кругов) по отображениям их граничных окружностей и т.д. В общем случае под $k$-мерной клеткой понимается $k$-мерный шар, и он приклеивается к $(k-1)$-мерному комплексу по отображению сферы на его крае. Обозначим через $c_i(K)$ – число клеток размерности $i$. Эйлерова характеристика $\chi(K)$ комплекса $K$ по определению равна $\sum_{i=0}^{\infty}(-1)^i c_i(K)$. Эйлерова характеристика является инвариантом, то есть не зависит от выбора разбиения комплекса $K$ на клетки. Отметим также, что на любом топологическом многообразии $M$ размерности $\dim(M) \leq 3$ можно ввести как гладкую, так и кусочно-линейную структуру. В книге [14] предлагается следующая интерпретация описанной выше ситуации в размерности $3$: любое трехмерное многообразие или любая кривая или поверхность в трехмерном многообразии кусочно-линейны, т.е. имеют комбинаторные триангуляции, но эти триангуляции выбраны настолько мелкими, что симплексы не различимы невооруженным глазом и многообразие, кривая или поверхность кажутся гладкими.

Напомним основное свойство сферического $s$-угольника: сумма его внутренних углов $(\alpha_1 + \alpha_2 + \alpha_3 + \ldots + \alpha_s)$ всегда больше чем $(s-2)\cdot 180°$. Разность $(\alpha_1 + \alpha_2 + \alpha_3 + \ldots + \alpha_s) - (s-2)\cdot\pi = \delta$, называется сферическим избытком, или сферическим эксцессом сферического $s$-угольника.

Если спроектировать $E$ ребер правильного многогранника $\{s, t\}$ (где $\{s, t\}$ – символ Шлефли многогранника) из центра на концентрическую сферу единичного радиуса, то они перейдут в $E$ дуг больших кругов, которые разобьют поверхность сферы на $F$ областей, "сферических $s$-угольников", где $t$ – число многоугольников сходящихся в одной вершине ($t$-гранный угол). Таким образом, многогранники порождают сферические мозаики (т.е. разбиения сферы) [15].

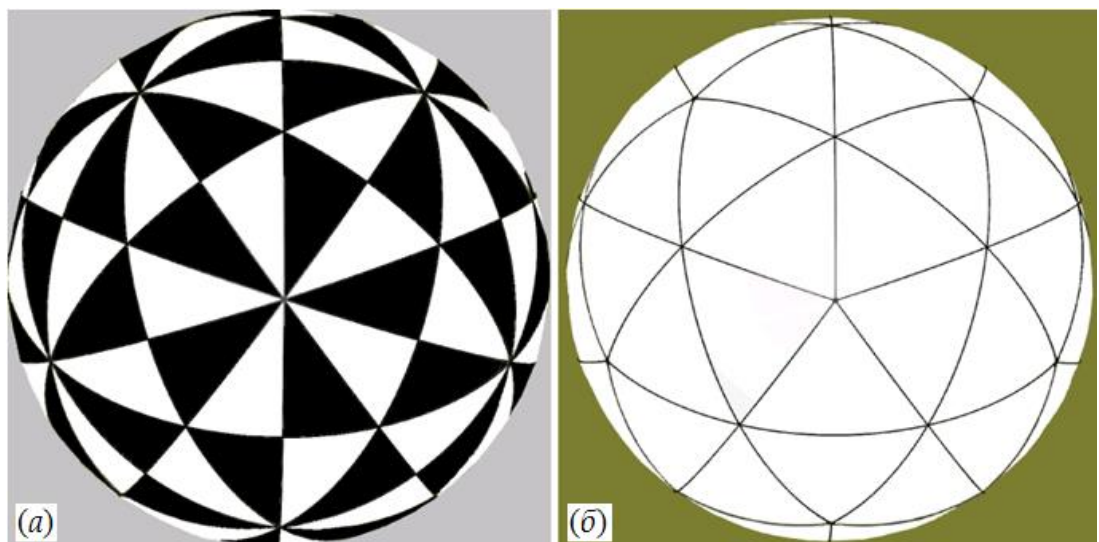

Рис. 1. Изоэдральные (не изогональные) разбиения сферы:

*а* – модель сферического гекзакисикосаэдра;

*б* – модель сферического пентакисдодекаэдра.

На рис. 1, *а* приведена модель сферического гекзакисикосаэдра, 120 одинаковых сферических треугольников с углами при вершинах (90°, 60°, 36°) образуют нормальное разбиение сферы, обладающее полной икосаэдральной группой симметрии, порядок которой равен $|I_\mathrm{h}| = 5! = 120$. Группа симметрии многогранника изображенного на рис. 1, *а* образуется из группы симметрии грани добавлением симметрии относительно сторон этой грани. Таким образом, ее порождают симметрии относительно трех сторон сферического треугольника Мёбиуса с углами $(\pi/5,\ \pi/3,\ \pi/2)$ символ которого (5, 3, 2). Все плоскости симметрии многогранника, а именно, плоскости, соединяющие середины противоположных ребер многогранника $\{s, t\}$ и дуального многогранника $\{t, s\}$, разбивают сферу на сеть таких треугольников, заполняющую всю поверхность сферы без зазоров и/или наложений. Этот сферический треугольник (площадь которого равна 6 квадратных единиц) является фундаментальной областью полной группы симметрий икосаэдра $I_h$, так как три образующие симметрии переводят его в соседние области.

Напомним, что площадь сферы единичного радиуса равна 720 квадратных единиц. Результатом деления 720/6 является число 120. Число 120 соответствует общему числу сферических треугольников на рис. 1, *а*.

Сферические треугольники на рис. 1, *а* поочередно закрашены черным и белым цветом (энантиоморфная раскраска), так что можно визуализировать и группу $I_h$, которая действием переводит любой треугольник во все остальные, и ее подгруппу вращений $I$ изоморфную знакопеременной группе перестановок $|A_5| = 5!/2 = 60$, которая действием переводит треугольник только в треугольник того же цвета.

Очевидно, что десять треугольников в центре рис. 1, *а* образуют грань сферического пентагонального додекаэдра, а шесть треугольников, сходящихся при одной вершине, образуют грань сферического икосаэдра. Заметим, что 24 склеенных треугольника образуют сферический двуугольник (разделив $120/24 = 5$, т.е. получим пятиугольный осоэдр). Относительное расположение полюсов на сфере (в данном случае) можно выбрать шестью различными способами.

Разбиение сферы, приведенное на рис. 1, *б* получается из разбиения рис. 1, *а* если склеить определенным образом каждые два смежных энантиоморфных треугольника (черного и белого цвета). Результирующий сферический треугольник (площадь которого 12 кв. ед.) является фундаментальной областью подгруппы вращений $I$ группы симметрии икосаэдра $I_h$. Укажем, что у разбиения сферы на рис. 1, *а* следующие значения (V = 62; F = 120; E = 180); у разбиения же на рис. 1, *б* – (V = 32; F = 60; E = 90), где V, F и E соответственно число вершин, граней и ребер. Отметим, что 720/12 = 60, где число 60 – это общее число сферических треугольников на рис. 1, *б*.

Выполним аналогичные расчеты для Платонова додекаэдра. У додекаэдра в каждой вершине сходится по три пентагональные грани (внутренний угол пентагона, как известно равен 108°). Умножив 108°·3 = 324°, далее 360° – 324° = 36°, потом 720/36 = 20, в результате число 20 – число вершин в додекаэдре. У сферического додекаэдра, разумеется, также пятиугольные грани, однако внутренний угол будет равен 120° (в каждой вершине сходится по три пятиугольника, т.е. 360°/3 = 120°). В результате умножив 120°·5 = 600°, далее 600° – 540° = 60° (число 540° получается так: 540° = 180°·(5 – 2)), потом 720/60 = 12, где число 12 – число граней в додекаэдре. И наоборот, поделив 720 на число граней (любого правильного многогранника) получим сферический избыток.

В случае же Архимедовых тел, например, усеченного икосаэдра, у которого 32 грани (20 гексагонов и 12 пентагонов) имеем, что 720/32 = 22.5, т.е. результат деления – рациональное число. Однако если выполнять деление 720 по отдельности на 12 и 20 результатом будут натуральные числа.

С анализом ограничения кристаллов и простых форм связано принятое в классической кристаллографии деление кристаллических классов в пределах каждой сингонии на голоэдрическую (т.е. "полногранную"), гемиэдрическую (имеющую половинное число граней от полного их числа в голоэдрическом классе), тетартоэдрическую (имеющую четверть граней) [16]. Голоэдрический класс – высший класс в данной сингонии, и его подгруппой является любой другой его класс. На рис. 1 также можно визуализировать гемиэдрию в голоэдрическом икосаэдральном классе, т.е. получаем из модели сферического гекзакисикосаэдра путем склеивания характерных совокупностей смежных треугольников (т.е. путем выделения подклассов в соответствующем классе) модели сферических: икосаэдра, додекаэдра, ромбического триаконтаэдра (тетартоэдрия), триакисикосаэдра и пентакисдодекаэдра (гемиэдрия). Сам же сферический гекзакисикосаэдр, изображенный на рис. 1, *а* можно получить, объединив пять сферических октаэдров соответственно ориентированных.

Для лучшего понимания дальнейших рассуждений, введем понятия протомногогранника и тривиального многогранника.

**Определение 6.** Протомногогранником называется многогранник, из которого путем симметрического склеивания определенных смежных граней (гемиэдрия) получается многогранник с вдвое меньшим числом граней. При этом склеивании точечные группы симметрии таких многогранников изоморфны.

Повторять такое склеивание граней можно до тех пор, пока не придем к тривиальному многограннику.

**Определение 7.** Тривиальным многогранником назовем многогранник, у которого любое симметрическое склеивание граней (гемиэдрия) приведет к изменению его точечной группы симметрии.

Изоэдральным протомногогранником с группой симметрии $I_h$ является гекзакисикосаэдр, а изоэдральным тривиальным многогранником с этой же группой будет пентагональный додекаэдр.

Если взять симметричный кристаллический многогранник и спроецировать на сферу не только его ребра (т.е. его скелет, каркас) но и "связи" соединяющие ближайшие "атомы" друг с другом, то получим разбиение сферы, группа симметрии которого изоморфна группе симметрии данного тела. Разбиения сферы, приведенные в данном исследовании, как раз получены по такому рецепту.

На рис. 2, *а* изображена модель сферического икосододекаэдра с дополнительно проведенными его плоскостями симметрии. Построение данных разбиений сферы подобно алгоритму Мёбиуса, который описан выше. Разбиение сферы, приведенное на рис. 2, *а* имеет следующие значения (V = 122; F = 240; E = 360).

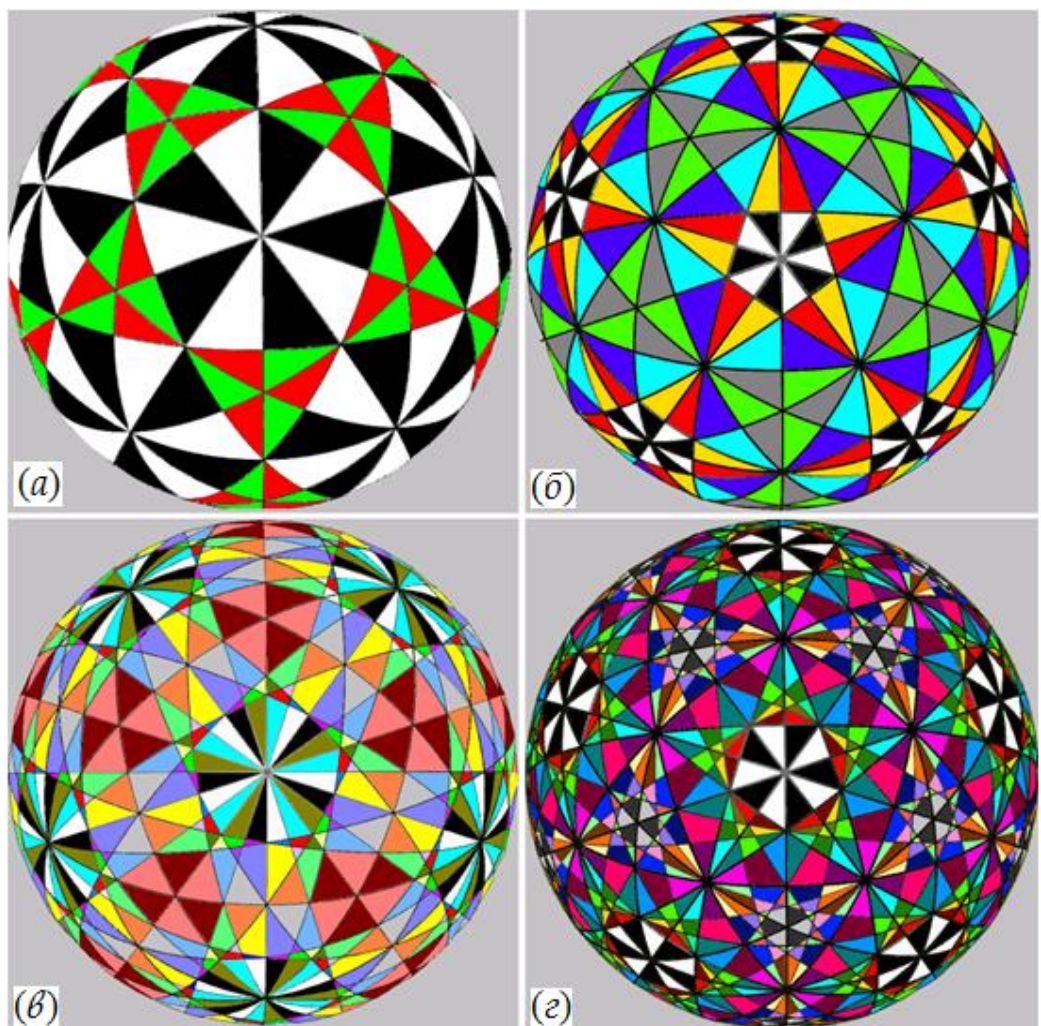

Рис. 2. Разбиения сферы, обладающие точечной группой симметрии $I_h$: в разбиении *а* участвует два вида неконгруэнтных сферических треугольников (N = 2); *б* – 4 вида треугольников (N = 4); *в* – 10 различных тайлов разбиения (N = 10); *г* – 18 различных плиток (N = 18).

Заметим, что в данном разбиении имеется два вида сферических треугольников. Такое разбиение сферы можно также получить, если объединить (комбинировать) соответственно ориентированные соосные сферические икосододекаэдр и гекзакисикосаэдр. Места пересечения геодезических линий можно считать вершинами модели и, следовательно, они могут быть декорированы атомами. Если на рис. 2, *а* объединить четыре треугольника (все разного цвета) сходящихся при одной вершине в одну фигуру, то получим сферический дельтоид. В совокупности данные дельтоиды образуют модель сферического дельтоидального гексеконтаэдра.

На рис. 2, *а, б и г* можно визуально установить, что данные разбиения являются некоторыми из вариантов разбиений икосододекаэдра, что позволяет говорить об обобщенной триангуляции Мёбиуса. На рис. 2, *а, б и г* число неконгруэнтных многоугольников участвующих в разбиениях сферических треугольников и пентагонов (т.е. граней сферического икосододекаэдра, которые можно визуализировать) соответственно равны: (1; 1), (1; 3), (8; 10).

Дуальное разбиение к разбиению, приведенному на рис. 2, *а* будет состоять из следующих сферических многоугольников: 12 декагонов; 20 гексагонов; 30 полуправильных восьмиугольников; 60 прямоугольников. В таком разбиении в каждой вершине сходится ровно по три ребра (однако, в соответствующем многограннике с плоскими гранями тригоноэдры не будут равными между собой). Несложно установить аналогичные характеристики сферических многогранников дуальных к многогранникам, изображенным на рис. 2, *б-г*. Заметим, что число различных тайлов в дуальных разбиениях к разбиениям, приведенным на рис. 1, *а* и рис. 2 будет на два больше. Другими словами, если например, в разбиении рис. 2, *а* участвует два неконгруэнтных треугольника, то в дуальном разбиении будет четыре различных многоугольника (рассмотренных выше). Отметим, что в дуальных разбиениях к разбиениям рис. 2, *б* и *в* в вершинах будет сходиться как по три, так и по четыре ребра, в следствии наличия у прямых разбиений (т.е. разбиений изображенных на рис. 2, *б* и *в*) как треугольников так четырехугольников.

Обратим внимание на то, что если соединить дугами окружностей точки на сфере полученные при решении одной из задач кристаллографии, а именно задачи о построении гномостереографической проекции кристалла, то мы реализуем сферическую модель дуального кристалла к данному.

Табл. 1. Данные о числе вершин, граней и ребер в разбиениях типа Пенроуза двумерной сферы обладающих точечной группой $I_h$.

| N  | V−2  | (V−2)/60 | F    | F/60 | F/120 | E    | E/60 | E/5 |
|----|------|----------|------|------|-------|------|------|-----|
| 1  | 60   | 1        | 120  | 2    | 1     | 180  | 3    | 36  |
| 2  | 120  | 2        | 240  | 4    | 2     | 360  | 6    | 72  |
| 3  | 180  | 3        | 360  | 6    | 3     | 540  | 9    | 108 |
| 4  | 240  | 4        | 480  | 8    | 4     | 720  | 12   | 144 |
| — |
| 9  | 660  | 11       | 1080 | 18   | 9     | 1740 | 29   | 348 |
| 10 | 780  | 13       | 1200 | 20   | 10    | 1980 | 33   | 396 |
| — |
| 14 | 1080 | 18       | 1680 | 28   | 14    | 2760 | 46   | 552 |
| — |
| 18 | 1380 | 23       | 2160 | 36   | 18    | 3540 | 59   | 708 |
| 19 | 1500 | 25       | 2280 | 38   | 19    | 3780 | 63   | 756 |
| ⋮ |

В табл. 1 приведены для конкретных разбиений сферы следующие данные: N – число различных тайлов; (V – 2) – число вершин (без двух); F – число граней; E – число ребер (т.е. число дуг на окружности которые образуют пары смежных сферических многоугольников). Отметим, что в столбце F/120 т.е. отношение числа граней в разбиении к порядку группы симметрии икосаэдра ($|I_h| = 120$) соответствует числу N в конкретном разбиении. В совокупности все N неконгруэнтных сферических многоугольников образуют разбиение фундаментальной области группы $I_h$ (т.е. разбиение сферического треугольника с углами (36°, 60°, 90°) – грани многогранника изображенного на рис. 1, *а*). Далее, можно заметить, что выполняется следующее равенство (V–2)/60 + F/60 = E/60, где число 60 – порядок циклической подгруппы группы $I_h$. В столбце E/5 можно для некоторых приведенных значений усмотреть следующую зависимость $180·n/5 = E/5$, где *n* – натуральное число.

В табл. 1 пустые строки означают, что разбиение, скажем с N = 5, не удалось получить в рамках алгоритма применяемого в данной работе. Список в табл. 1 продолжается, и далее, скажем можно внести данные и о разбиениях рассмотренных ниже (см. рис. 3). Число аналогичного рода различных вариантов мозаик на сфере велико. Очевидно, что существуют разбиения сферы с числом F → ∞.

Если удастся поместить в каждой вершине разбиения сферы (с конечным числом F) атомы, причем определенного сорта, то гипотетически такие сферические кристаллы (молекулы) могут быть экспериментально реализованы. Например, для этого одним из инструментов может служить "квантовый пинцет".

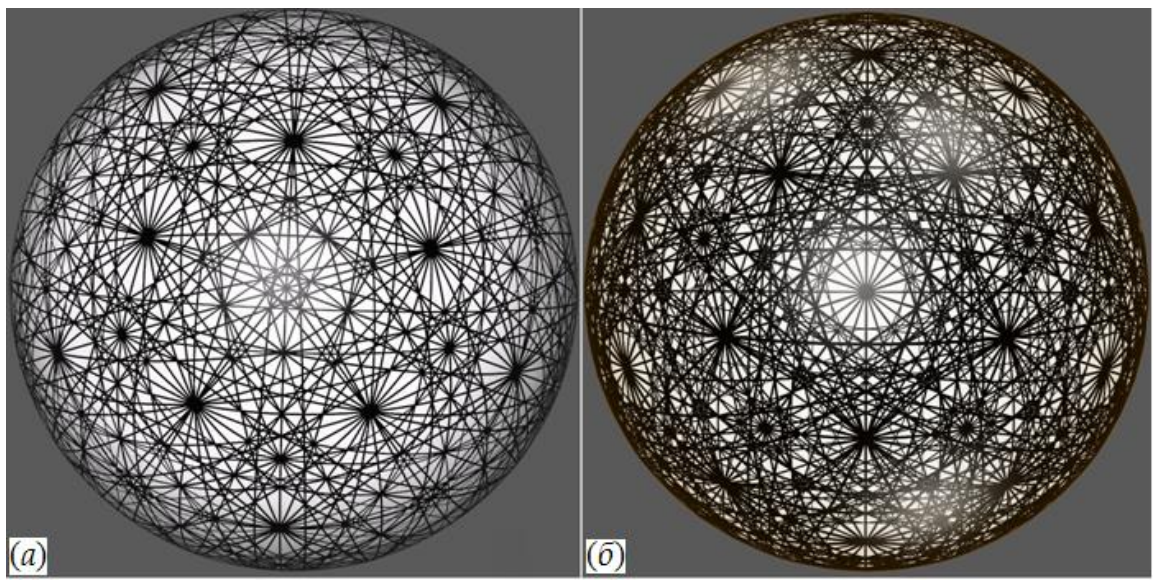

Рис. 3. Построение предельного ряда разбиений сферы типа Пенроуза обладающих икосаэдральной группой симметрии $I_h$; общее число сферических многоугольников участвующих в разбиении *а* меньше чем на *б*.

При дальнейшем построении предельного ряда реализовываются такие разбиения сферы, что 2-клетки становятся визуально неразрешимыми. В пределе, когда F → ∞ получаем серию сферических бесконечногранников. Для изучения таких моделей необходимо увеличение разрешительной способности, т.е. радиус сферы нужно устремить также к ∞. Геометрия на сфере локально евклидова (т.е. гауссова кривизна сферы равна $+1,$ но локально кривизна равна нулю). Из этого следует, что при построении предельного ряда на сфере, когда $F \to \infty$ можно локально исследовать (в окрестности осей симметрии) фрагменты квазипериодических разбиений евклидовой плоскости, обладающих симметрией пятого порядка.

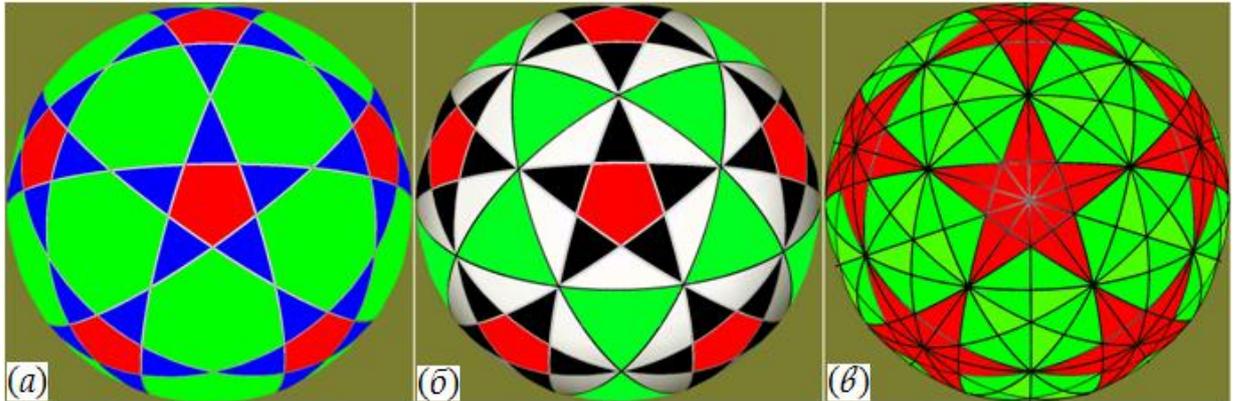

Рис. 4. Сферическая модель почти многогранника Джонсона – полностью усечённого усечённого икосаэдра (*а*) и его производные многогранники (*б, в*).

В геометрии почти многогранником Джонсона называется выпуклый многогранник, в котором грани близки к правильным многоугольникам, однако некоторые (или все) из них не совсем правильные. Напомним, что многогранник называется многогранником Джонсона, если он не является ни Платоновым телом, ни архимедовым, ни каталановым, ни призмой, ни антипризмой. На рис. 4, *а* изображена модель полностью усечённого усечённого икосаэдра. Сферический многогранник на рис. 4, *а* является дуальным многогранником к сферическому ромбическому девяностограннику. На рис. 4, *а* и *б* соответственно следующие значения $(V = 90; \ F = 92; \ E = 180)$ и $(V = 90; \ F = 152; \ E = 240).$ На рис. 4, *б* и *в* приведены производные сферические многогранники, полученные из многогранника, изображенного на рис. 4, *а*. Многогранник на рис. 4, *в* подобен многограннику, приведенному на рис. 2, *б*, однако по-иному раскрашен. Таким образом, проводя плоскости симметрии у почти многогранника Джонсона можно также получать новые модели сферических многогранников. Другими словами из тривиального многогранника производя определенным образом построение геодезических линий можно получать ряд производных многогранников. В пределе получим серию бесконечногранных протомногогранников.

Очевидно, что разные пути склейки граней одного протомногогранника могут привести к различным тривиальным многогранникам.

Отметим, что сферу можно рассматривать как конечный клеточный комплекс, клетками которого являются точки (вершины), дуги (ребра), 2-клетки – многоугольники. Предположим, что в вершинах клеточного разбиения расположены атомы разного сорта. Тогда совокупность данных атомов составят кристаллическое множество атомов сферической геометрии (определение кристаллического множества атомов геометрии приведено в книге [8]).

Рассмотрим пример построения кристаллического множества атомов на сфере, которые получаются посредством проекции атомов составляющих тело ромбического триаконтаэдра и с последующим выполнением соответствующих разбиений сферы.

Проекции на сферу атомов приведенных на рис. 5 выполнялись посредством следующего преобразования их координат

$$\xi_1 = \frac{x}{\sqrt{x^2+y^2+z^2}}, \ \xi_2 = \frac{y}{\sqrt{x^2+y^2+z^2}}, \ \xi_3 = \frac{z}{\sqrt{x^2+y^2+z^2}},$$

где $x$, $y$, $z$ – координаты атомов в кристалле ромбического триаконтаэдра, а $\xi_1$, $\xi_2$, $\xi_3$ – координаты атомов на поверхности сферы. Если соединить данные точки на сфере (рис. 5, *а* и *в*) отрезками то получим кусочно-гладкую поверхность, варианты которой изображены на рис. 5, *б* и *г*. Далее можно спроецировать на сферу данные отрезки прямых, и получим серию разбиений сферы, которые будут отличны от разбиений рассмотренных выше (на рис. 2-4 можно заметить целые окружности, а на разбиений рис. 5 – только дуги окружностей).

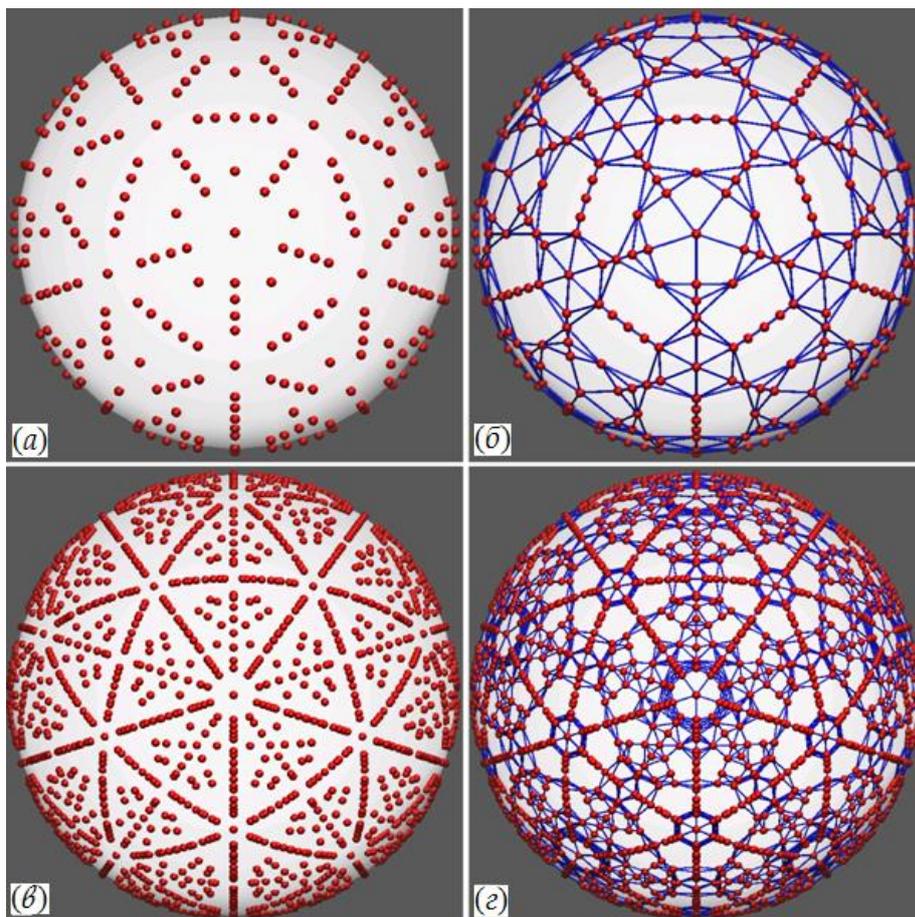

Рис. 5. Проекция на сферу кристаллических множеств атомов, из которых состояли модели непустых ромбических триаконтаэдров (тел) и соответствующие разбиения сферы:

*а* – модель, состоящая из 729 атомов; *б* – модель из 4096 атомов.

Следует отметить, что если спроецировать вершины разбиений надлежаще ориентированных сфер на стереографическую сетку полярного или экваториального типа (в том числе и комбинированную сетку Флинта), то это в свою очередь позволит решить ряд задач кристаллографии связанных с геометрическими характеристиками квазикристаллов.

Тот в книге [17] пишет: упомянем, например, следующую интересную, но довольно трудную задачу. Пусть на сфере задано $m$ материальных точек, снабженных одинаковыми электрическими зарядами; полагая, что эти точки могут свободно двигаться по сфере, требуется найти расположение точек, отвечающих их устойчивому равновесию, т.е. такое расположение, при котором собственный потенциал точек достигает абсолютного минимума. Еще не доказано даже, что решение этой задачи при $m = 6, 12$ и $\infty$ дается вершинами правильных многогранников $\{3, 4\}$, $\{3, 5\}$ и $\{3, 6$ или $\infty\}$. Очевидно, что если на сфере имеется некоторое число атомов, то данные атомы будут стремиться занять такие положения, в которых система будет иметь минимум энергии. Причем, если данное число атомов (соответствующих сортов) совпадет с числом N (установленными в данной работе), то равновесное расположение атомов будет совпадать с точками пересечения геодезических (т.е. вершинами сферических многогранников). Число сортов атомов сопоставимо с числом неконгруэнтных тайлов в конкретном разбиении поверхности.

В дальнейшем представляется интересным применить идеи Шубникова [18], а также Белова и Тархова [19] о антисимметрии и цветной симметрии к анализу предельного ряда разбиений типа Пенроуза.

Перейдем теперь к рассмотрению одного из важных вопросов комбинаторной геометрии, касающегося нахождению хроматического числа пространства. Напомним, что хроматическим числом евклидовой плоскости называется минимальное число цветов $\phi$, для которого существует такая раскраска точек плоскости в один из цветов, что никакие две точки одного цвета не находятся на расстоянии ровно 1 друг от друга. Задача Нельсона–Эрдёша–Хадвигера ставит вопрос о минимальном числе цветов, в которые можно раскрасить $n$-мерное евклидовое пространство. Та же задача имеет смысл для произвольного метрического пространства. Отметим работу А. М. Райгородского [20] в которой поставлена задача и приведен ряд ярких и важных результатов связанных с нахождение хроматического числа евклидового пространства.

В пространстве $E^2$, хроматическое число по оценкам находится в приделах $4 \leq \phi \leq 7$ и, насколько мне известно, равно 7, причем продвинуться дальше до сих пор не удаётся. Примером раскраски евклидовой плоскости в 7 цветов является разбиение ее грань в грань равными гексагонами семи разных цветов. Если раскрасить плоскость разбитую грань в грань на квадраты, то число $\phi$ будет равным 9. Однако плоскость, разбитую на квадраты, но не грань в грань, а в виде "кирпичной кладки", то $\phi$ будет равным 7. Введем определение координационного числа многогранника.

**Определение 8.** Если пространство выполнено конгруэнтными многогранниками, то координационным числом к многогранника назовем число многогранников контактирующих с данным.

Следовательно, для евклидовой плоскости хроматическое число можно рассчитать: $\phi = \kappa + 1$. Так как для случая разбиения плоскости $E^2$ на гексагоны $\kappa = 6$, то $\phi = 6 + 1 = 7$.

По аналогии пространство $E^3$ разобьем на кубы (с диаметром сферы в которую вписан куб немного меньше 1). Так как каждый куб окружен 26 соседними кубами то ($\kappa = 26$), следовательно, в пространстве $E^3$ хроматическое число $\phi \leq 27$. По аналогии с двумерным случаем, рассмотрим выполнения пространства $E^3$ конгруэнтными кубами (или гексагональными призмами) также в виде "кирпичной кладки". Очевидно $\kappa = 14$, следовательно, получим, что в пространстве $E^3$ верхняя граница хроматического числа $\phi = 15$.

В рамках данного исследования представляет интерес рассмотрение вопроса о хроматическом числе пространства $S^2$. Рассмотрим две модели изоэдральных тривиальных многогранников: сферического додекаэдра и сферического тетраэдра (см. рис. 6).

Двенадцать граней додекаэдра (см. рис. 6, *а*) раскрашены шестью различными светами. Причем каждая пара граней, через которые проходит одна из шести поворотных осей пятого порядка, раскрашена одним цветом. У сферического пентагонального додекаэдра κ = 5, следовательно, в пространстве $S^2$ хроматическое число ϕ ≤ 6. На рис. 6, *б* приведена модель сферического тетраэдра, у которого κ = 3. Таким образом, в пространстве $S^2$ верхняя граница хроматического числа ϕ ≤ 4. Очевидно, что нижняя граница 2 ≤ ϕ. При этом возникает вопрос: может ли на двумерной сфере ϕ = 2 (если эту сферу разделить на две полусферы)?

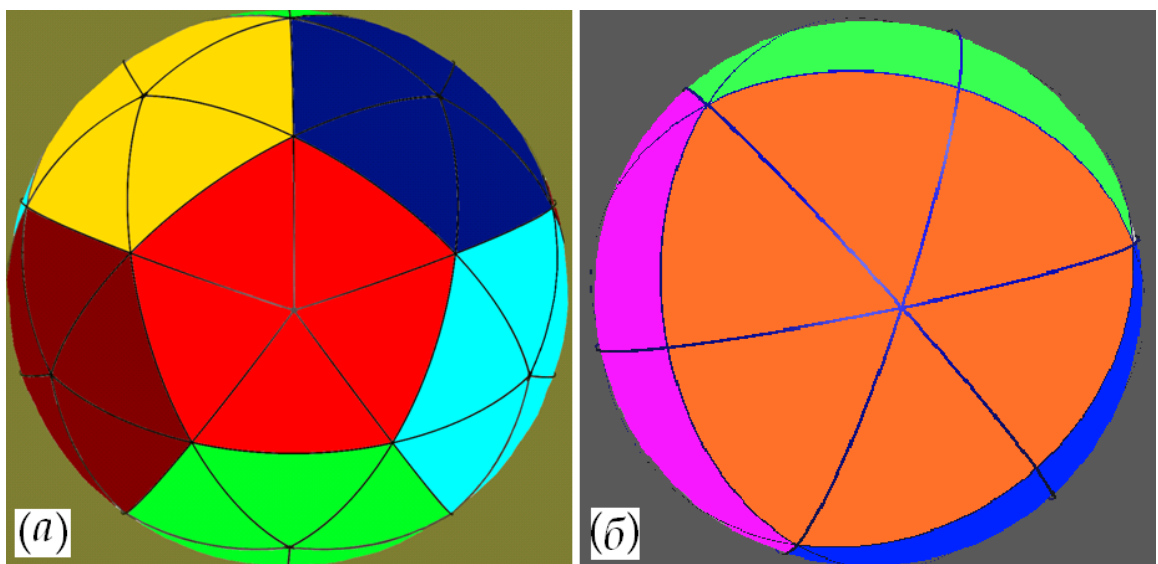

Рис. 6. Цветная раскраска сферического додекаэдра (*а*) и тетраэдра (*б*).

Отметим, что для измерения углового расстояния между двумя точками на сфере можно использовать, например, сетку Вульфа. Если же нужно узнать абсолютное значение расстояния между двумя точка на сфере, то необходимо соответствующее угловое расстояние разделить на длину экватора.

Представляется интересным поставить задачу о нахождении хроматического числа в геометриях с римановой метрикой: $H^2$, $S^2$, $H^3$, $S^3$, $S^2{\times}R$, $H^2{\times}R$, $SL_2R$, Nil, Sol.

Напоследок приведем одно из знаменитых высказываний Анри Пуанкаре [21] «… как же теперь отнесемся мы, после всего сказанного, к вопросу: истинна ли евклидова геометрия? Вопрос этот является бессмысленным. С равным успехом можно бы задать вопрос …: верны ли декартовы координаты и ложны ли полярные? Одна геометрическая система не может быть вернее другой; она может быть лишь более удобной».

### 3. Выводы

В работе решена задача о построении предельного ряда разбиений типа Пенроуза двумерной сферы.

Число различных неконгруэнтных многоугольников в конкретном квазипериодическом разбиении сферы можно сопоставить с числом разных сортов атомов в соответствующем ему реальном сферическом кристалле (молекуле). Атомы в реальном кристалле будут стремиться расположиться в вершинах пересечения геодезических линий сферического кристалла моделирующего данный.

Пользуясь моделью сферического 120-гранника (гекзакисикосаэдра – изоэдрального протомногогранника) и комбинаций его с другими многогранниками, описан рецепт построения икосаэдра, додекаэдра, тел Каталана, а также других многогранников, обладающих, в том числе полной икосаэдральной группой симметрий $I_h$. Аналогичные реализации сферических многогранников Платона, Архимеда и Каталана, обладающих полной октаэдральной группой симметрий $O_h$ можно произвести, используя модель сферического гекзакисоктаэдра, а для тетраэдральной симметрии $T_h$ – модель сферического триакистетраэдра.

Исходя из принципа эквивалентности Ламберта между геометрией Римана и геометрией Евклида, приведен алгоритм расчета численных характеристик многогранников.

Установлена верхняя граница значения хроматического числа в пространстве $S^2$, которая равна 4.

Подобного рода модели многогранников также могут соответствовать проекциям на сферу кристаллов со сложной скульптурой граней, в том числе в пределе и бесконечногранников.

# 4. Список литературы

1. Делоне Б. Н. Теория планигонов. Изв. АН СССР серия математическая. 23 (1959). С. 365-386.

2. Ле Ты Куок Тханг, Пиунихин С. А., Садов В. А. // Успехи Математических наук. 1993. Т. 48. Вып. 1 (289). С. 41.

3. Федоров Е. С. Симметрия на плоскости. – Зап. Мин. Общ., 2-я серия, 1891, т. XXVIII, стр. 345-390.

4. Шубников А. В. К вопросу о строении кристаллов Изв. АН. Сер.6, 1916г. Т.10, № 9. С. 755-779.

5. Федоров Е. С. Начала учения о фигурах. М.: Изд-во Академии наук СССР, 1953. 404 с.

6. Шубников А. В. Избранные труды по кристаллографии М., «Наука», 1975. с. 556.

7. Дышлис А. А., Плахтиенко Н. П. Модели нанокристаллов и неклассические периодические функции. LAP: Lambert Academic. Publishing., Deutschland, 2014. 303 с.

8. Дышлис А. А., Покась С. М. Геометрия Лобачевского и ее применение в математике и кристаллографии. Издательство LAP LAMBERT Academic Publishing RU. 2017. 689 с.

9. Laves F. Ebenenteilung und Koordinationszahl. Zeitschr. für Kristallogr, Bd. 78 (1930) P. 208-241.

10. Мадисон А. Е. Симметрия квазикристаллов. Физика твердого тела, 2013, том 55, вып.4. С. 784-796.

11. Клейн Ф. В сб.: Об основаниях геометрии. Сборник классических работ по геометрии Лобачевского и развитию ее идей. / Под ред. А. П. Нордена. ГИТТЛ, М. (1956). 399 с.

12. Wenninger M. Polyhedron Models. Cambridge. Cambridge University Press. 1974. 208 p.

13. Schwarz H. A. / Journal für die reine und angewandte Mathematik. 1873. T. 75. C. 292-335.

14. Матвеев С. В., Фоменко А. Т. Алгоритмические и компьютерные методы в трехмерной топологии. – М.: Изд-во МГУ, 1991. 301 с.

15. Кокстер Г. С. М. Введение в геометрию. Пер. с англ. А. Б. Катка и С.Б. Катом под ред. Б. А. Розенфельда и И. М. Яглома. М., Наука. 1966. 648 с.

16. Вайнштейн Б. К. Современная кристаллография (в четырех томах). Том 1. Симметрия кристаллов. Методы структурной кристаллографии. М.: Наука, 1979. 384 с.

17. Тот Л. Ф. Расположения на плоскости, на сфере и в пространстве. Перевод с немецкого Н. М. Макаровой, Редакция перевода, приложения и примечания И. М. Яглома. Государственное издательство физико-математической литературы. Москва, 1958. 365 с.

18. Шубников А. В. Симметрия и антисимметрия конечных фигур. Изд-во АН СССР, 1951. С. 7.

19. Белов Н. В., Тархова Т. Н. // Кристаллография. 1956. Т. 1. Вып. 1, 4. С. 618.

20. Райгородский А. М. Хроматические числа. Серия: «Библиотека "Математическое просвещение"». Вып. 28. М.: МЦНМО, 2003. 44с.: ил.

21. Пуанкаре А. Наука и гипотеза. СПб, 1906. 58 с.